\documentclass[10pt,conference,a4paper]{IEEEtran}
\IEEEoverridecommandlockouts
\usepackage{cite}
\usepackage{amsmath,amssymb,amsfonts}
\usepackage{algorithmic}
\usepackage{graphicx}
\usepackage{textcomp}
\usepackage{xcolor}
\def\BibTeX{{\rm B\kern-.05em{\sc i\kern-.025em b}\kern-.08em
    T\kern-.1667em\lower.7ex\hbox{E}\kern-.125emX}}

\usepackage{url}

\usepackage{graphicx}
\usepackage{diagbox}

\usepackage{xspace}
\usepackage{xcolor} %
\usepackage{tabularx}
\usepackage{subfig}
\usepackage{enumitem}
\usepackage[T1]{fontenc}
\usepackage[utf8]{inputenc}
\usepackage{xurl}
\usepackage{amsmath}
\usepackage{amssymb}
\usepackage{tikz}
\usepackage{soul}
\usepackage{multirow}
\usepackage{booktabs} %
\usepackage{csquotes} %
\usepackage[normalem]{ulem} %
\usepackage[capitalize,noabbrev]{cleveref}

\usepackage{listings}
\usepackage{siunitx} 
\usepackage{pdfpages}
\usepackage{pdflscape}
\usepackage{pgfplotstable}
\usepackage{pgfplots}
\usepgfplotslibrary{colorbrewer}
\usetikzlibrary{patterns}

\usepackage{etoolbox}
\preto\subequations{\ifhmode\unskip\fi}

\usepackage{float}

\renewcommand{\paragraph}[1]{\noindent\textbf{#1.}}

\newcounter{lineno}

\renewcommand{\paragraph}{\textbf} 
    \def\@IEEEsectpunct{.\ }

    \def\parax{\@startsection%
        {paragraph}%
        {4}%
        {0\parindent}%
        {0.6ex plus 0.1ex minus 0.1ex}%
        {0ex}%
        {\normalfont\normalsize\bfseries}%
        *%
    }%
\makeatother

\newcommand{\para}[1]{\parax{#1}\ } %

\newcommand{\examp}[1]{\parax{\emph{#1}}\ } %

\definecolor{rbettercolor}{HTML}{AAFFAA}
\definecolor{rworsecolor}{HTML}{FFAAAA}

\definecolor{Cfree}{HTML}{3bac87}

\definecolor{Cpartlyfree}{HTML}{b9a350}

\definecolor{Cnotfree}{HTML}{9868a0}

\pgfplotsset{compat=1.16}

\begin{document}

\title{Charting Censorship Resilience and Global Internet Reachability: A Quantitative Approach\\}

\author{\IEEEauthorblockN{Marina Ivanović}
  \textit{ETH Zurich}\\
  \and
  \IEEEauthorblockN{François Wirz}
  \textit{ETH Zurich}\\
  \and
  \IEEEauthorblockN{Jordi Subirà Nieto}
  \textit{ETH Zurich}\\
  \and
  \IEEEauthorblockN{Adrian Perrig}
  \textit{ETH Zurich}\\
}

\maketitle

\IEEEpubid{\begin{minipage}{
    \textwidth}\ \\[12pt]
  ISBN 978-3-903176-63-8~\copyright~2024 IFIP
\end{minipage}}

\IEEEpubidadjcol

\begin{abstract}

    Internet censorship and global Internet reachability are prevalent topics of today's Internet. Nonetheless, the impact of network topology and Internet architecture to these aspects of the Internet is under-explored. With the goal of informing policy discussions with an objective basis, we present an approach for evaluating both censorship resilience and global Internet reachability using quantitative network metrics, which are applicable to current BGP/IP networks and also to alternative Internet network architectures. We devise and instantiate the metric on the network topology of multiple countries, comparing the BGP/IP network, an overlay network using a waypoint mechanism for circumventing undesired nodes, and the path-aware Internet architecture SCION. The novelty of the approach resides in providing a metric enabling the analysis of these aspects of the Internet at the routing level, taking into account the innate properties of the routing protocol and architecture. We demonstrate that the Internet topology matters, and strongly influences both censorship resilience and reachability to the global Internet. Finally, we argue that access to multiple paths accompanied with path-awareness could enable a higher level of censorship resilience compared to the current Internet, and reduce the centralization of Internet routing.

\end{abstract}

        
\section{Introduction}
\label{sec:intro}

The issue of Internet censorship---the deliberate restriction or suppression of
information~\cite{Hall2023, Leberknight2010}---has emerged as a pervasive concern in the digital
era.
There have been long standing records of censored network
communication practices employed by various entities, and most prominently
governments~\cite{Aryan2013, Ramesh2020a, Singh2020a, Marczak2015a, Gill2015a}. Furthermore, the issue of dependency on certain countries has also grown in the context of the global Internet~\cite{Karlin2009, Shah2016}, with various analyses that western countries have gained significant influence on the global Internet routing, as routing paths predominantly traverse them~\cite{Edmundson2018}.

The innate properties of Internet topologies and architectures play a crucial role in determining how traffic flows through the network, and
therefore, they are likely to influence the effectiveness of censorship efforts, and in general Internet reachability. In the context of the global Internet, we refer to the network topology as the interconnectedness of
Autonomous Systems (ASes), while the Internet architecture encompasses the underlying structure and
protocols that facilitate operation of the Internet.
For instance, the core Internet routing protocol is the Border Gateway Protocol (BGP), which provides Internet inter-domain routing~\cite{rfc4271BGP}.
In traditional networks using BGP, ASes only consider the next hop when making routing decisions.
Unlike traditional routing, SCION---a next-generation Internet architecture designed
to provide secure inter-domain routing~\cite{Chuat2022}---ensures that packets traverse
predetermined paths and making end-nodes in the network path-aware~\cite{Chuat2022}. Finally, the usage of Virtual Private Network (VPNs) has been a popular technique for
Internet censorship evasion, given that it could not only provide an additional layer of secrecy using
encryption, but also circumvent censoring devices altogether\cite{Tschantz2016a,Khattak2016a}.

Previous research underlines the evidence that the topology of the network could be an indicator of
deployed censorship capabilities~\cite{Leyba2019, Roberts2011a, Salamatian2021}, and reachability to the global Internet ~\cite{Shah2016, Edmundson2018, Acharya2017}. Nonetheless, to the best of our knowledge, it is an open research challenge how traditional BGP routing, the use of waypoint network with VPN nodes, and in general fundamentally different approaches such as BGP and SCION could be quantitatively compared in this context.

\para{Research Question}
In this context, the following research questions arise:
\emph{Do the topology and the architecture of the Internet have an influence on Internet censorship,
and in general global Internet reachability?}
And if so, \emph{can we quantify this influence?}
Answering these questions does not only provide insights into the interplay between Internet
topology and architecture, and censorship and reachability, but can also quantitatively inform policy-makers.
To achieve this, we propose a concrete approach for evaluating censorship and reachability
aspects using a quantitative network metric.

\para{Key Contributions}

\begin{enumerate}
  \item We design a quantitative metric instantiable to \emph{censorship resilience}
	  and \textit{global Internet reachability}.
	  The metric is agnostic to network topology, and applicable to the current Internet and captures path-awareness.
(Section~\ref{sec:metrics}).
  \item We instantiate the metric on the current Internet topology of several countries, analyzing their network topologies with regards to Internet censorship. In the context of the influence to Internet reachability, we instantiate our metric using diverse groups of potentially influential countries (Section~\ref{sec:eval}).
  \item We perform extensive experiments using the contemporary Internet topology on both BGP, a waypoint network with intermediate nodes, and SCION, a
	  path-aware Internet architectur, providing a comparative
	  analysis for Internet censorship and global reachability (Section~\ref{sec:results}).
\end{enumerate}

\section{Background}
\label{sec:background}

\para{Internet Censorship} Internet censorship can be observed when an entity in power restricts its citizens or users from certain online communication or content, if it is deemed harmful, politically inappropriate, sensitive or legally noncompliant~\cite{Hall2023, Leberknight2010}.
Censorship could happen at various communication points: at the \emph{end-point devices}, or \emph{on-link}, by nodes that the traffic passes through~\cite{Khattak2016a}.
Prior work indicates that the traversed nodes and the network topology do play a role for censorship~\cite{Roberts2011a, Acharya2017, Wrana2024}, stressing the need for a quantitative evaluation of this influence.
To that end, the scope and the focus of this article will be on the latter, \emph{on-link} censorship.

\para{Censorship Circumvention} In the face of widely deployed censorship techniques, a plethora of censorship circumvention methods have arisen~\cite{Tschantz2016a}. Among others, one can use intermediate nodes in the network as a waypoint, to circumvent the censors' influence~\cite{Khattak2016a}, for instance by utilizing a Virtual Private Network (VPN) connection.

\para{Global Internet Reachability} Prior work shows a certain hegemonic influence on global Internet reachability~\cite{Karlin2009}, while many depend on Western countries to access common Internet destinations~~\cite{Edmundson2018}.
Not only can this jeopardize global reachability due to dependence on other countries and their Internet infrastructure~\cite{Edmundson2018}, but it can also raise concerns about potential surveillance~\cite{Edmundson2016, Edmundson2016b, Obar2013} and collateral damage~\cite{Acharya2017}.

\para{SCION Next-generation Internet Architecture} SCION is a \emph{next-generation Internet architecture}~\cite{Chuat2022, Bechtold2014}, which is already deployed in production networks~\cite{Kreuhenbeuhl2021}.
SCION groups ASes into Isolation Domains (ISDs), with shared
governance institutions. An ISD provides a trust environment between the ASes in it, which could---among others---group around a common jurisdiction~\cite{Chuat2022}.
SCION is a multi-path and path-aware architecture~\cite{Trammell2018}.
In general, each node can have several end-to-end paths at its disposal, from which it can select any one of them. This contrasts with the BGP/IP Internet that is hop-by-hop routed.

\section{Quantifying Avoidability in a Network}
\label{sec:metrics}

In this section, we introduce \emph{Avoidability Potential}, drawing from classic graph theory metrics like group betweenness centrality.

\subsection{Avoidability Potential}

\textit{Avoidability Potential} quantifies the potential of avoiding undesirable or potentially malicious nodes in the network.
In this work, we focus on the topological organization of the Internet,
providing an analysis at the inter-AS routing level.

\para{Network Model} We model the network as a graph $\mathcal{G}(\mathcal{V}, \mathcal{E})$.
Each node $v \in \mathcal{V}$ represents an Autonomous System (AS), whereas each edge $e \in
\mathcal{E}$ represents a link between nodes.
Edges between nodes are labeled with the standard business relationships on the Internet:
\emph{customer-provider} and \emph{peer-peer}~\cite{Gao2001}.
Furthermore, the nodes in the SCION network are grouped into ISDs. This does not affect the model of
the network as a graph, but rather only provides grouping of nodes composing it.

\para{Threat Model} In a graph $\mathcal{G}(\mathcal{V}, \mathcal{E})$, communication between nodes $s, d \in \mathcal{V}$ can be censored, intercepted, blocked,  or in any way
tampered with. This interference can occur anywhere on their path due to unreliable, potentially malicious, simply or untrusted nodes. We assume that ASes pose a threat as a whole, with varying numbers of Byzantine ASes that may collude. While the motives and interests for Internet censorship are diverse, our model does not explicitly address them.

\para{The Metric} We define set $\mathcal{S} \subset \mathcal{V}$ as the set of all
\emph{source} nodes from which paths of interest originate, and the set $\mathcal{D} \subset
\mathcal{V}$ as the set of all \emph{destinations}, where the paths terminate.
Finally, we define $\mathcal{X} \subset \mathcal{V}$ as a set of ASes that should be avoided when
communicating between nodes of interest.

Given the graph $\mathcal{G}(\mathcal{V}, \mathcal{E})$, the set $\mathcal{X}$ of nodes whose avoidability is
analyzed, a source node $s \in \mathcal{S}$ and a destination node $d \in
\mathcal{D}$, we define $e_{\mathcal{X}}(s \rightarrow d)$ as a binary flag of whether a path between $s$ and $d$ exists, which completely circumvents
nodes in $\mathcal{X}$.
If it exists, we say that these nodes have the full potential of establishing a connection.

\[
  e_{\mathcal{X}}(s \rightarrow d)= 
  \begin{cases}
      1,& \exists r \text{, a path } s \rightarrow d \text{, s.t. } \forall x \in \mathcal{X}, x \notin r \\
      0,& \text{otherwise}
\end{cases}
\]

From there, we define the \emph{Avoidability Potential} by allowing for all possible sources $s \in
\mathcal{S}$ and destinations $d \in \mathcal{D}$.
This yields the final metric, presented in the Equation~\eqref{eq:avoid_pot}.

\begin{equation}
  AP_\mathcal{X}(\mathcal{S}, \mathcal{D}) = \frac{\sum\limits_{\substack{s \in \mathcal{S} \\ d \in \mathcal{D}}} e_{X}(s \rightarrow d)}{\|\mathcal{S}\| \cdot \|\mathcal{D}\|}
  \label{eq:avoid_pot}
\end{equation}

The value $\|S\| \cdot \|D\|$ in the Equation \eqref{eq:avoid_pot} is the number of all pairs of sources and destinations, which leads to a normalized value $AP_\mathcal{X}(\mathcal{S},
\mathcal{D}) \in [0, 1]$.
Here, $1$ means that the nodes from $\mathcal{D}$ can always receive traffic from the nodes in
$\mathcal{S}$, without traversing any node in $\mathcal{X}$, whereas $0$ would mean that this
traffic would \emph{always} traverse some of these nodes.

The above introduced metric is general, applicable to any graph and sets of nodes in the graph
$\mathcal{S}$, $\mathcal{D}$ and $\mathcal{X}$.
It is also independent of the network model, routing protocol, and captures architectures which allow for multiple paths.
Thereupon we lay out two important applications of this metric, which are largely relevant for
today's Internet: censorship resilience, and global Internet reachability.

\subsection{Censorship Resilience Potential}

We apply the \textit{Avoidability Potential} to the case of censorship resilience, deriving the \textit{Censorship Resilience Potential} metric, where the set $\mathcal{X}$ is the set of censoring nodes
$\mathcal{C}$.

\begin{equation}
  CRP_\mathcal{C}(\mathcal{S}, \mathcal{D}) = AP_\mathcal{C}(\mathcal{S}, \mathcal{D})
  \label{eq:crp_is_ap}
\end{equation}

\examp{Example: National Outflow Traffic}
We can focus on a specific country by defining $\mathcal{C}$ as the set of ASes with interests or capabilities to censor outflow traffic. For outflow traffic originating within the country and heading to a foreign AS, all national ASes are sources, thus part of set $\mathcal{S}$. Similarly, ASes outside the country are in the destination set $\mathcal{D}$.

\para{Towards a Metric Agnostic to Normative Claims} As mentioned briefly in Section~\ref{sec:background}, censorship occurs globally for diverse reasons, including political motives. While one might consider that quantifying censorship must incorporate these interests and provide normative justifications, we present a metric that views all network nodes as potential censors without delving into individual motives or activities\footnote{We do however note that our metric can also be applied to a set of censoring ASes that are
\emph{a priori} labeled as such.}.

\para{Defining Censoring ASes} In the case where censoring ASes are not known \emph{a priori}, we
define them based on their potential to choke the highest number of paths from $\mathcal{S}$
to $\mathcal{D}$.
Following the intuition of the outflow traffic
---an example
relevant for the current Internet~\cite{Xu2011a,Leyba2019}---
we define the set of censors $\mathcal{C}$ as a subset of $\mathcal{S}$, which have the
\emph{highest potential} of choking outflow paths that go from
$\mathcal{S}$ to $\mathcal{D}$.
In other words, we focus on the situation where $\mathcal{S} \cap \mathcal{D} = \emptyset$ and
$\mathcal{C} \subset \mathcal{S}$. 

A border AS is an AS in $\mathcal{S}$, with at least one direct link to an AS outside of $\mathcal{S}$~\cite{Leyba2019}.
Set $\mathcal{B}$ is the set of all border ASes.

\begin{equation}
  \mathcal{B} = \{b \in \mathcal{S} | \exists e = (b, x) \in E \text{ s.t. } x \notin \mathcal{S}\}
  \label{eq:border_as}
\end{equation}

Following the work by Leyba et al.~\cite{Leyba2019}, we adapt the concept of choke potential to capture the concept of path-awareness.
For that, consider a subset of border ASes, $\mathcal{B}' \subset \mathcal{B}$.
Their \emph{Cumulative Choke Potential} ($CPP$) is the fraction of outflow paths that they could
choke together.
The rigorous definition of $CCP_{\mathcal{S}, \mathcal{D}}(\mathcal{B}')$ is shown in Equation
\ref{eq:chokepot}, and due to normalization yields to $CCP_{\mathcal{S},
\mathcal{D}}(\mathcal{B}') \in [0, 1]$.

\begin{equation}
  CCP_{\mathcal{S}, \mathcal{D}}(\mathcal{B}') = \frac{\sum\limits_{\substack{s \in \mathcal{S} \\ d \in \mathcal{D}}} f_{\mathcal{B}'}(s \rightarrow d) }{\|\mathcal{S}\| \cdot \|\mathcal{D}\|}
  \label{eq:chokepot}
\end{equation}

\[
  f_{\mathcal{B}'}(s \rightarrow d)= 
  \begin{cases}
      1,& \forall r \text{, a path } s \rightarrow d, \exists b' \in \mathcal{B}' \textit{, s.t. } b' \in r \\
      0,& \text{otherwise}
\end{cases}
\]

We define the set of censoring ASes $\mathcal{C}$ as a subset of $\mathcal{B}$ with a cardinality of $\|\mathcal{C}\| = N$, collectively capable of choking the highest number of outflow paths. As an intuition, if all border ASes were to censor, their cumulative choke potential would be 1, resulting in $CRP_\mathcal{C}(\mathcal{S}, \mathcal{D}) = 0$. However, strict enforcement of censorship across all border ASes is challenging. Thus, we conduct experiments across various countries with different values of $N$, elaborated on in Section~\ref{sec:eval}.

\para{Algorithm for \emph{CRP} Metric} In a graph $\mathcal{G}(\mathcal{V}, \mathcal{E})$, one must define the sets of source and destination nodes, $\mathcal{S} \subset \mathcal{V}$ and $\mathcal{D} \subset \mathcal{V}$, respectively. These sets can be chosen arbitrarily or based on node properties, such as country of origin. The set of censoring ASes $\mathcal{C} \subset \mathcal{V}$ can be defined in two ways.

\begin{figure}[htbp] %
  \centering
  \includegraphics[width=\linewidth]{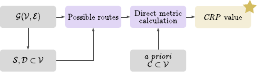}
  \caption{\emph{Censorship Resilience Potential} ($CRP$): the algorithm pipeline, with censoring ASes known \emph{a priori}.}
  \label{fig:pipeline_crp}
\end{figure}

In the first method, $\mathcal{C}$ is known \emph{a priori}. The metric's value is determined by calculating the fraction of paths from $\mathcal{S}$ to $\mathcal{D}$ that avoid censoring ASes. The algorithm pipeline for this method is depicted in Figure~\ref{fig:pipeline_crp}, with graph and set inputs marked in gray, intermediate steps in purple, and the outputs in yellow.

\begin{figure}[htbp] %
  \centering
  \includegraphics[width=\linewidth]{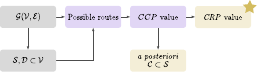}
  \caption{\emph{Censorship Resilience Potential} ($CRP$): the algorithm pipeline, where the  \emph{Cumulative Choke Potential} ($CCP$) is used for defining the set of
	censoring ASes $\mathcal{C}$ \emph{a posteriori}.}
  \label{fig:pipeline_crp_ccp}
\end{figure}

The second method determines the value of $CRP$ using \emph{Cumulative Choke Potential} ($CCP$).
In this case, the underlying assumption is that censoring ASes would be border ASes from the set
$\mathcal{S}$.
The $CCP$ value of the subset of them provides us with both the set of censoring ASes  $\mathcal{C}$
with high potential of cumulatively choking outflow traffic, and the value of the final metric.
The full algorithm pipeline of this method is laid out on Figure~\ref{fig:pipeline_crp_ccp}, with
the same color-coding from Figure~\ref{fig:pipeline_crp}.

\para{\emph{CRP} as Means for Comparative Analysis} Our metric aims to facilitate comparative analysis of different network models and Internet architectures. If the set of censoring ASes is known beforehand, the method outlined in Figure~\ref{fig:pipeline_crp} should be applied universally. However, if the set $\mathcal{C}$ is not predetermined, it should be defined independently. For a comprehensive analysis, $\mathcal{C}$ should be defined according to the pipeline depicted in Figure~\ref{fig:pipeline_crp_ccp} for each architecture. We utilize this approach in our simulation, discussed further in Section~\ref{sec:eval}, yielding results suitable for comparing BGP, waypoint models, and SCION.

\subsection{Global Reachability Potential}
\label{sec:global_reachability}

Our second goal is to gauge the potential of reaching the global Internet, while avoiding
undesirable nodes in the network.
We achieve this independently of the specific network topology, or the Internet architecture, by
adopting the \textit{Avoidability Potential} metric to this use case.

\para{The Metric} Given the graph representing the global Internet,
$\mathcal{G}(\mathcal{V}, \mathcal{E})$, it is possible that certain nodes are more central to for global connectivity than
others.
To measure how much influence a group of nodes $\mathcal{X} \subset \mathcal{V}$ has on nodes
$\mathcal{S} = \mathcal{V} \setminus \mathcal{X}$ to establish paths with each other, we employ the
\textit{Avoidability Potential} metric, for convenience calling it \textit{Global Reachability
Potential}.

\begin{equation}
  GRP_\mathcal{X}(\mathcal{S}, \mathcal{S}) = AP_\mathcal{X}(\mathcal{S}, \mathcal{S}), \text{   } \mathcal{S} = \mathcal{V} \setminus \mathcal{X}
  \label{eq:grp_is_ap}
\end{equation}

\examp{Example: Collateral Damage of Internet Censorship} While censorship techniques primarily target specific network nodes~\cite{Gill2015a}, they can also lead to collateral damage affecting other nodes beyond the intended scope~\cite{Sparks2012a}. For instance, Acharya et al. suggest that countries known for censorship may impact global reachability~\cite{Acharya2017}. Our metric can analyze this collateral censorship damage at the AS level, extending its applicability to such scenarios.

\examp{Example: Influence of Hegemonic Groups} 
Several authors note that a small number of ASes serve as global transit networks~\cite{Fontugne2017}, raising concerns about their hegemonic influence on global Internet reachability~\cite{Karlin2009}. The \emph{Global Reachability Potential} metric can analyze the potential for circumventing such influential nodes, offering comparative analysis of network models and quantitative evidence of Internet routing centralization or democratization.

\section{Experimental Evaluation}
\label{sec:eval}

In this section, we outline the setup and approach for extensive experiments, whose results we elaborate on in Section~\ref{sec:results}.

\para{Overview of Analyzed Network Models}
In our study, we apply our metrics to three different network models.
First, we look at the current BGP/IP Internet design, and without
considering any routing attacks.
Second, we consider a scenario where waypoint mechanisms are widely used to bypass undesirable nodes in the network.
Lastly, we examine SCION, a path-aware next-generation Internet architecture, where each end-host can choose the whole end-to-end path.

\subsection{Datasets}
\label{sec:datasets}

For all our experiments, we use the datasets that represents real and contemporary relationships between ASes.

\para{AS Relationships} We utilize the CAIDA AS Relationships dataset~\cite{CAIDA_as_relationships_serial_2} to model the network graph $\mathcal{G}(\mathcal{V}, \mathcal{E})$. This dataset offers the Internet's topology, employed directly in all BGP and waypoint model simulations. We maintain consistency in our SCION simulation by employing the same topology, ensuring comparable and relevant results.

\para{AS Country Origin} 
To accurately determine the country origin of each AS, we employ two datasets. First, the CAIDA AS to Organizations Mappings offers legal entity country origins for ASes~\cite{CAIDA_as_org_info}. Second, the RIPEStat Geo Map dataset~\cite{RIPE_Stat} provides physical locations where ASes announce BGP prefixes. We account for Tier-1 ASes being present in multiple countries, interconnected through branches.

\para{Waypoints in the Network} For the waypoint network model, we utilize the anonymous dataset from MaxMind. This dataset identifies ASes previously associated with potential host anonymization services, such as VPN or Tor nodes~\cite{MaxMind}.

\subsection{Country Network}

We apply the \emph{Censorship Resilience Potential} metric to network nodes according to their country of origin. Our selection of countries encompasses diversity based on various indicators, such as geographical location, population size, national network size, and the \emph{Internet Freedom Score} (IFS)~\cite{FreedomHouseInternetFreedom}.

\para{Nodes Forming a Country Network}  Let $\mathfrak{X}$ denote a country of interest, and $\mathcal{K}$ represent the set of ASes originating from $\mathfrak{X}$. We define $\mathcal{K}$ as follows:

\begin{equation}
    \mathcal{K} = \{k \in \mathcal{V} | \text{ country}({k}) = \mathfrak{X}\} 
    \label{eq:country_origin}
\end{equation}

We use set $\mathcal{K}$ to define a country's network, excluding outliers. Inspired by the study of Guillermo et al. on the global Internet\cite{Guillermo2022}, we identify \emph{islands} in a country network, labeling the largest connected component as the country network.

\subsection{SCION Topology}
\label{sec:scion:topo}

Although SCION is already deployed, its current production network footprint does not yet reach the scale of the BGP
infrastructure.
To address this, we construct the SCION topology using the CAIDA AS Relationship dataset~\cite{CAIDA_as_relationships_serial_2}, ensuring it reflects real-world deployment and remains comparable to the BGP/IP network.

\para{Core ASes} We define core ASes as per the analysis by Krähenbühl et al.~\cite{Kreuhenbeuhl2021}, based on the customer cone size of ASes. When determining the value of the \emph{Global Reachability Potential}, we use the graph of core ASes, as discussed in the remainder of this paper.

\para{Grouping into ISDs} Grouping nodes into an ISD is essential for applying the \textit{Censorship Resilience Potential} metric, especially on a per-country basis. We assume that ASes connected to a country's network infrastructure naturally form a connected component, creating a ``national'' ISD---or a group of ISDs in the general case~\cite{Chuat2022}. We keep links between all ASes in an ISD and disregard links to ASes outside the ISD.

\subsection{Simulation on Diverse Network Models}
In this section we comment on the implementation details.

\para{Censoring ASes} We conduct inter-AS BGP simulation using routing tree algorithm by Gill et al. to determine preferred paths between ASes in the Internet topology~\cite{Gill2012}. In the BGP network model, we define $\mathcal{C = C_{BGP}}$ as a subset of border ASes with the highest potential to restrict outflow traffic. Drawing on Leyba et al.'s work, we attribute potentially choked paths to the last border ASes~\cite{Leyba2019}.
Additionally, since the waypoint model shares the same topology and routing algorithm as the BGP model, we employ the same set of censoring ASes for consistency.
Finally, for the SCION network topology, we select a subset of border ASes with the highest customer cone size, forming the set of censoring nodes, denoted as $\mathcal{C = C_{SCION}}$. This method is an effective heuristic for selecting nodes with the highest potential to cumulatively choke outflow paths, as it considers the customer cone size of each border AS in the country network.

\para{Censorship Resilience Potential} Once the set of censoring ASes $\mathcal{C}$ is established, we compute the \emph{Censorship Resilience Potential} metric for both the BGP and waypoint models by assessing the fraction of paths not intercepted by nodes in $\mathcal{C}$. In SCION, sources have the freedom to choose the entire end-to-end path. Thus, we determine whether exists a path that can leave the ISD while avoiding nodes in $\mathcal{C}$.

\para{Global Reachability Potential} Once the set of nodes analyzed for global influence $\mathcal{X}$ is defined, we compute the \emph{Global Reachability Potential} metric for both the BGP and waypoint models by assessing the fraction of paths not intercepted by nodes in $\mathcal{X}$. For SCION, it is enough analyzing the interconnectedness of the core ASes, as they are crucial for the global Internet reachability.

\section{Experimental Results}
\label{sec:results}

We apply our metric to two distinct cases: censorship resilience of various countries, and global Internet reachability, commenting on their results in this section. For extended results refer to Section Availability.

\subsection{National Censorship Resilience Potential}

\begin{figure*}[htbp] %
  \centering
  \includegraphics[width=\linewidth]{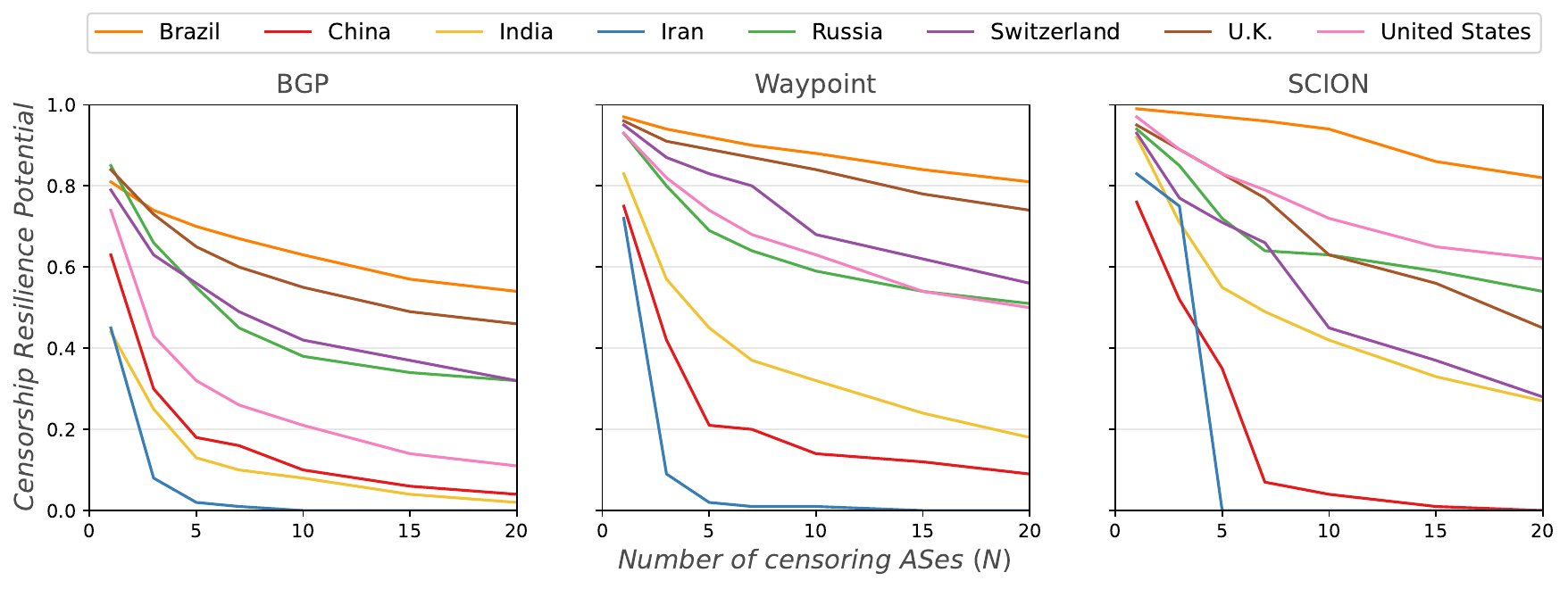}
  \caption{\emph{Censorship Resilience Potential} for BGP, waypoint model, and SCION, presented for various countries and varying number of censoring ASes, $N$.}
  \label{fig:graphic_results}
\end{figure*}

We assess the \emph{Censorship Resilience Potential} metric across BGP, waypoint model, and SCION, incorporating diverse countries. Our findings underscore the critical influence of network topology on Internet censorship. Figure~\ref{fig:graphic_results} presents our results, which we analyze further in this section.

\para{Border ASes as Central Choking Points} Our findings reveal that even a small number of a country's border ASes can significantly restrict outflow paths in the current Internet. With as few as 20 border ASes, up to 50\% of outflow paths can be choked, irrespective of the country network's size. For instance, in BGP, the United States could potentially choke 26\% of outflow paths ($CRP = 0.74$) with just one AS. We complement this analysis with network statistics in Table~\ref{tab:country_stats}, indicating network size and the number of border ASes across all models. For example, comparing the Iranian and Swiss BGP-network topologies, we observe Iran's more centralized model with fewer border ASes, corroborating its centralized censorship model~\cite{Aryan2013, Salamatian2021}. Conversely, Switzerland exhibits a higher density of border ASes. This comparison offers insights into other countries' censorship efforts and the potential collaboration among ASes in such endeavors.

\newcommand{\ra}[1]{\renewcommand{\arraystretch}{#1}}
\begin{table}
  \centering
  \ra{1.3}
  \begin{tabular}{@{}rrrr@{}}
    \toprule
    & Number & Border ASes & Border ASes \\
    & of ASes & (BGP/Waypoint) & (SCION) \\
    \midrule
    Brazil & 8174 & 2285 (28\%)  & 243 (3\%) \\ 
    China & 534 & 94 (18\%)  & 21 (4\%) \\ 
    India & 2537 & 209 (8\%)  & 36 (1\%) \\ 
    Iran & 481 & 25 (5\%)  & 5 (1\%) \\ 
    Russia & 4957 & 1139 (23\%)  & 82 (2\%) \\ 
    Switzerland & 654 & 308 (47\%)  & 28 (4\%) \\ 
    U.K. & 1562 & 861 (55\%)  & 62 (4\%) \\  
    United States & 17934 & 2173 (12\%)  & 236 (1\%) \\ 
    \bottomrule
  \end{tabular}
  \caption{Country network statistics: number of total and border ASes in a country network, across all analyzed models.}
  \label{tab:country_stats}
\end{table}

\para{Network Topology Matters} The results confirm that the number of border ASes is not the sole influential factor. The interconnectedness of ASes within the network also matters, regardless of the Internet architecture. In other words, even with numerous exit points from a country network, the routing among them might not be evenly distributed, leading to dependency on a small number of ASes. For instance, the United States, with nearly 18'000 ASes, exhibits a relatively low CRP value of 0.32, attributed to only 5 ASes and their routing influence.

\para{Path-selection and Censorship Resilience} Multiple exit paths from a country offer potential for enhancing censorship resilience, yet they are often underutilized. In path-aware technologies like the waypoint model, the likelihood of a single AS controlling all outflow paths is noticeably reduced compared to BGP.
In SCION, the lower number of Border ASes, due to technical and governance factors, results in fewer exit points in the network (see Table~\ref{tab:country_stats}). However, our analysis demonstrates that path-selection mechanisms can significantly mitigate the impact of censoring Border ASes by offering alternatives to bypass undesired ASes.

\subsection{Global Internet Reachability}

We assess the \emph{Global Reachability Potential} metric outlined in section \ref{sec:global_reachability} across BGP, waypoint model, and SCION. The selected influential country groups align with previous studies~\cite{Karlin2009, Fontugne2017}, identifying nations with either negative impact on Internet reachability or potential for collateral damage due to censorship. Results are summarized in Table~\ref{tab:global_reachability_results}.

\para{Nodes Centrality} The results indicate that ASes from the United States and the Five Eyes countries serve as transit nodes for over 40\% of all paths between nodes in different countries. Similarly, ASes from the European Union play a central role in global reachability. We also assess the \emph{Global Reachability Potential} of Iran, China, and Russia, commonly discussed in the context of censorship~\cite{Aryan2013, Ensafi2015a, Ramesh2020a} and its potential collateral damage~\cite{Acharya2017}. Our analysis reveals that they do not exert significant influence on global reachability across the three network models analyzed, as they lack central positioning in the current Internet topology.

\begin{table}[tb]
  \centering
  \ra{1.3}
  \begin{tabular}{@{}rrrr@{}}
    \toprule
    & BGP & Waypoint & SCION \\
    \midrule
    United States & 0.59 & 0.92 & 0.9951 \\
    Five Eyes & 0.52 & 0.88 & 0.9941 \\
    European Union & 0.87 & 0.98 & 0.9975 \\
    Iran, China, Russia & 0.98 & 0.99 & 0.9995 \\
    \bottomrule
  \end{tabular}
  \caption{Results of the \textit{Global Reachability Potential}, with various groups of countries analyzed for global influence.}
  \label{tab:global_reachability_results}
\end{table}

\para{Path-selection and Global Reachability} Results in Table~\ref{tab:global_reachability_results} reveal variations among the three analyzed models. The impact of examined groups on global Internet reachability is lower in the waypoint model compared to BGP. Multiple waypoint hosts in this model suggest the existence of various paths, increasing the potential to bypass undesired ones. However, as several waypoint nodes originate from the United States, achieving full \emph{Global Reachability Potential} remains challenging. Additionally, SCION's end-to-end path-awareness offers means to circumvent undesirable nodes, resulting in high \textit{Global Reachability Potential} for all analyzed country groups.

\section{Related Work}
\label{sec:related}

\para{Country Network Analysis} In the realm of Internet censorship, prior studies by various authors~\cite{Master2023a} emphasize the significance of network topology. For instance, Ensafi et al. note that Tor traffic, often censored in China, bypasses censorship when entering via CERNET~\cite{Ensafi2015a}. Gill et al. characterize Iran's network as centralized~\cite{Gill2015a}, while Salamatian et al. highlight the limited direct links of Iran's network to foreign ASes, suggesting strategic use of BGP for censorship~\cite{Salamatian2021}. Additionally, Wählisch et al. employ a sector-based approach to assess betweenness centrality in the German network~\cite{Waehlisch2012}.

\para{Control of National Outflow Traffic} 
Roberts et al. developed a measure of network complexity, unveiling underlying properties indicative of a country's censorship capabilities~\cite{Roberts2011a}. Similarly, Leyba et al. found that the number of nodes capable of choking a significant fraction of outflow paths is not only low but also decreasing over time~\cite{Leyba2019}.

\para{Global Internet Reachability} Other researchers examined global Internet reachability by assessing betweenness centrality on a global scale, pinpointing nodes and countries pivotal for global connectivity~\cite{Karlin2009, Fontugne2017}, and delving into the potential collateral damage from censorship efforts~\cite{Acharya2017}.

\para{Broadened Prior Work and Contributions}  Our metric builds on prior work to create a comprehensive tool for quantifying censorship resilience and global Internet reachability. It is versatile across various network models and topologies, including path-aware Internet architectures. Importantly, it does not require predefined censoring ASes for assessing censorship resilience.

\para{Next-generation Internet Architectures and Censorship} We underscore the significance of scrutinizing next-generation Internet architectures in the context of censorship and Internet reachability. While Kohler~\cite{Kohler2022} and Wrana et al.~\cite{Wrana2024} explored this qualitatively, our main contribution is a quantitative metric suitable for comparative analysis.
\section{Discussion}
\label{sec:discussion}

\para{Routing Attacks on BGP} When assessing \emph{Censorship Resilience Potential} and \emph{Global Reachability Potential} metrics on BGP and the waypoint model, we establish legitimate paths between any two nodes, without considering routing attacks by malicious actors. Such attacks could involve redirecting traffic~\cite{rfc4272BGPVulner}, jeopardizing both censorship resilience and global Internet accessibility.

\para{Waypoint Model on the Internet} Our waypoint model sheds light on the influence of systems like VPN connections in bypassing undesirable nodes. However, it might oversimplify censorship circumvention by suggesting reliance on waypoint service providers, whereas censoring entities often block IP addresses from such providers. Nevertheless, it offers quantitative evidence of the advantages of multiple paths in Internet routing.

\para{Internet Deployments} BGP serves as the exclusive inter-domain routing protocol, rendering our BGP results directly relevant to the current Internet. Conversely, SCION's limited deployment~\cite{Kreuhenbeuhl2021} positions our findings as a future prospect. Moreover, waypoints, such as VPN connections, can complement SCION since their deployment is independent and compatible.

\para{Policy Impacts} We avoid making normative claims about Internet censorship or global reachability. Instead, we provide an objective metric for their evaluation, serving as a quantitative tool. This approach can offer policymakers insights into network technology design, development, and deployment.

\section{Conclusion}
\label{sec:conclusion}

In this paper, we have demonstrated how network topology and Internet architecture can affect a country's resilience to Internet censorship and its global reachability, highlighting dependencies on specific network nodes. We have proposed a novel approach that utilizes quantitative network metrics to evaluate these aspects of today's Internet. We evaluated the metric across diverse Internet models using Border Gateway Protocol (BGP), a model of waypoints based on Virtual Private Networks (VPNs), and SCION path-aware architecture. Our results underscore the importance of network topology and suggest that path-aware architectures could democratize global routing, potentially enhancing censorship resilience.

\section*{Ethics Statement}

In this work we use datasets already publicly available, and do not conduct any additional Internet measurements that may raise ethical concerns.

\section*{Availability}
\label{sec:availability}

The accompanying code repository and extended results are available at: https://github.com/IvanovicM/charting-censorship

\bibliography{bibtex/bib/IEEEabrv.bib,bibtex/bib/references.bib}{}

\begin{thebibliography}{10}
\providecommand{\url}[1]{#1}
\csname url@samestyle\endcsname
\providecommand{\newblock}{\relax}
\providecommand{\bibinfo}[2]{#2}
\providecommand{\BIBentrySTDinterwordspacing}{\spaceskip=0pt\relax}
\providecommand{\BIBentryALTinterwordstretchfactor}{4}
\providecommand{\BIBentryALTinterwordspacing}{\spaceskip=\fontdimen2\font plus
\BIBentryALTinterwordstretchfactor\fontdimen3\font minus
  \fontdimen4\font\relax}
\providecommand{\BIBforeignlanguage}[2]{{%
\expandafter\ifx\csname l@#1\endcsname\relax
\typeout{** WARNING: IEEEtran.bst: No hyphenation pattern has been}%
\typeout{** loaded for the language `#1'. Using the pattern for}%
\typeout{** the default language instead.}%
\else
\language=\csname l@#1\endcsname
\fi
#2}}
\providecommand{\BIBdecl}{\relax}
\BIBdecl

\bibitem{Hall2023}
J.~L. Hall, M.~D. Aaron, A.~Andersdotter, B.~Jones, N.~Feamster, and M.~Knodel,
  ``{A Survey of Worldwide Censorship Techniques},'' Internet Engineering Task
  Force, Internet-Draft draft-irtf-pearg-censorship-09, Jan. 2023, work in
  Progress.

\bibitem{Leberknight2010}
C.~S. Leberknight, M.~Chiang, H.~V. Poor, and F.~Wong, ``A taxonomy of internet
  censorship and anticensorship,'' \emph{Fifth International Conference on Fun
  with Algorithms}, 2010.

\bibitem{Aryan2013}
S.~Aryan, H.~Aryan, and J.~A. Halderman, ``Internet censorship in iran: A first
  look,'' in \emph{3rd USENIX Workshop on Free and Open Communications on the
  Internet (FOCI 13)}.\hskip 1em plus 0.5em minus 0.4em\relax Washington, D.C.:
  USENIX Association, Aug. 2013.

\bibitem{Ramesh2020a}
R.~Ramesh, R.~S. Raman, M.~Bernhard, V.~Ongkowijaya, L.~Evdokimov,
  A.~Edmundson, S.~Sprecher, M.~Ikram, and R.~Ensafi, ``Decentralized control:
  A case study of {Russia},'' in \emph{Network and Distributed System
  Security}.\hskip 1em plus 0.5em minus 0.4em\relax The Internet Society, 2020.

\bibitem{Singh2020a}
K.~Singh, G.~Grover, and V.~Bansal, ``How {India} censors the web,'' in
  \emph{Web Science}.\hskip 1em plus 0.5em minus 0.4em\relax ACM, 2020.

\bibitem{Marczak2015a}
B.~Marczak, N.~Weaver, J.~Dalek, R.~Ensafi, D.~Fifield, S.~McKune, A.~Rey,
  J.~Scott-Railton, R.~Deibert, and V.~Paxson, ``An analysis of {China}'s
  ``{Great Cannon}'','' in \emph{Free and Open Communications on the
  Internet}.\hskip 1em plus 0.5em minus 0.4em\relax USENIX, 2015.

\bibitem{Gill2015a}
P.~Gill, M.~Crete-Nishihata, J.~Dalek, S.~Goldberg, A.~Senft, and G.~Wiseman,
  ``Characterizing web censorship worldwide: Another look at the {OpenNet
  Initiative} data,'' \emph{Transactions on the Web}, vol.~9, no.~1, 2015.

\bibitem{Karlin2009}
J.~Karlin, S.~Forrest, and J.~Rexford, ``Nation-state routing: Censorship,
  wiretapping, and {BGP},'' \emph{arXiv}, 2009.

\bibitem{Shah2016}
A.~Shah, R.~Fontugne, and C.~Papadopoulos, ``Towards characterizing
  international routing detours,'' in \emph{Proceedings of the 12th Asian
  Internet Engineering Conference}, ser. AINTEC '16.\hskip 1em plus 0.5em minus
  0.4em\relax New York, NY, USA: Association for Computing Machinery, 2016, p.
  17–24.

\bibitem{Edmundson2018}
A.~Edmundson, R.~Ensafi, N.~Feamster, and J.~Rexford, ``Nation-state hegemony
  in internet routing,'' in \emph{Proceedings of the 1st ACM SIGCAS Conference
  on Computing and Sustainable Societies}, ser. COMPASS '18.\hskip 1em plus
  0.5em minus 0.4em\relax New York, NY, USA: Association for Computing
  Machinery, 2018.

\bibitem{rfc4271BGP}
Y.~Rekhter, S.~Hares, and T.~Li, ``{A Border Gateway Protocol 4 (BGP-4)},'' RFC
  4271, Jan. 2006.

\bibitem{Chuat2022}
\BIBentryALTinterwordspacing
L.~Chuat, M.~Legner, D.~A. Basin, D.~Hausheer, S.~Hitz, P.~M{\"{u}}ller, and
  A.~Perrig, \emph{{The Complete Guide to {SCION} - From Design Principles to
  Formal Verification}}, ser. Information Security and Cryptography.\hskip 1em
  plus 0.5em minus 0.4em\relax Springer, 2022. [Online]. Available:
  \url{https://doi.org/10.1007/978-3-031-05288-0}
\BIBentrySTDinterwordspacing

\bibitem{Tschantz2016a}
M.~C. Tschantz, S.~Afroz, Anonymous, and V.~Paxson, ``{SoK}: Towards grounding
  censorship circumvention in empiricism,'' in \emph{Symposium on Security \&
  Privacy}.\hskip 1em plus 0.5em minus 0.4em\relax IEEE, 2016.

\bibitem{Khattak2016a}
S.~Khattak, T.~Elahi, L.~Simon, C.~M. Swanson, S.~J. Murdoch, and I.~Goldberg,
  ``{SoK}: Making sense of censorship resistance systems,'' \emph{Privacy
  Enhancing Technologies}, vol. 2016, no.~4, pp. 37--61, 2016.

\bibitem{Leyba2019}
K.~G. Leyba, B.~Edwards, C.~Freeman, J.~R. Crandall, and S.~Forrest, ``Borders
  and gateways: Measuring and analyzing national as chokepoints,'' in
  \emph{Proceedings of the 2nd ACM SIGCAS Conference on Computing and
  Sustainable Societies}, ser. COMPASS '19.\hskip 1em plus 0.5em minus
  0.4em\relax New York, NY, USA: Association for Computing Machinery, 2019, p.
  184–194.

\bibitem{Roberts2011a}
H.~Roberts, D.~Larochelle, R.~Faris, and J.~Palfrey, ``Mapping local {Internet}
  control,'' in \emph{Computer Communications Workshop}.\hskip 1em plus 0.5em
  minus 0.4em\relax IEEE, 2011.

\bibitem{Salamatian2021}
L.~Salamatian, F.~Douzet, K.~Salamatian, and K.~Limonier, ``{The geopolitics
  behind the routes data travel: a case study of Iran},'' \emph{Journal of
  Cybersecurity}, vol.~7, no.~1, p. tyab018, 08 2021.

\bibitem{Acharya2017}
H.~B. Acharya, S.~Chakravarty, and D.~Gosain, ``Few throats to choke: On the
  current structure of the internet,'' in \emph{2017 IEEE 42nd Conference on
  Local Computer Networks (LCN)}, 2017, pp. 339--346.

\bibitem{Wrana2024}
M.~Wrana, D.~Barradas, and N.~Asokan, ``The spectre of surveillance and
  censorship in future internet architectures,'' \emph{arXiv}, 2024.

\bibitem{Edmundson2016}
A.~Edmundson, R.~Ensafi, N.~Feamster, and J.~Rexford, ``A first look into
  transnational routing detours,'' in \emph{Proceedings of the 2016 ACM SIGCOMM
  Conference}, ser. SIGCOMM '16.\hskip 1em plus 0.5em minus 0.4em\relax New
  York, NY, USA: Association for Computing Machinery, 2016, p. 567–568.

\bibitem{Edmundson2016b}
\BIBentryALTinterwordspacing
------, ``Characterizing and avoiding routing detours through surveillance
  states,'' \emph{CoRR}, vol. abs/1605.07685, 2016. [Online]. Available:
  \url{http://arxiv.org/abs/1605.07685}
\BIBentrySTDinterwordspacing

\bibitem{Obar2013}
J.~Obar and A.~Clement, ``{Internet Surveillance and Boomerang Routing: A Call
  for Canadian Network Sovereignty},'' \emph{SSRN Electronic Journal}, 2013.

\bibitem{Bechtold2014}
S.~Bechtold and A.~Perrig, ``Accountability in future internet architectures,''
  \emph{Commun. ACM}, vol.~57, no.~9, p. 21–23, sep 2014.

\bibitem{Kreuhenbeuhl2021}
C.~Kr\"{a}henb\"{u}hl, S.~Tabaeiaghdaei, C.~Gloor, J.~Kwon, A.~Perrig,
  D.~Hausheer, and D.~Roos, ``Deployment and scalability of an inter-domain
  multi-path routing infrastructure,'' in \emph{Proceedings of the 17th
  International Conference on Emerging Networking EXperiments and
  Technologies}, ser. CoNEXT '21.\hskip 1em plus 0.5em minus 0.4em\relax New
  York, NY, USA: Association for Computing Machinery, 2021, p. 126–140.

\bibitem{Trammell2018}
B.~Trammell, J.-P. Smith, and A.~Perrig, ``Adding path awareness to the
  internet architecture,'' \emph{IEEE Internet Computing}, vol.~22, no.~2, pp.
  96--102, 2018.

\bibitem{Gao2001}
L.~Gao and J.~Rexford, ``Stable internet routing without global coordination,''
  \emph{IEEE/ACM Transactions on Networking}, vol.~9, no.~6, pp. 681--692,
  2001.

\bibitem{Xu2011a}
X.~Xu, Z.~M. Mao, and J.~A. Halderman, ``{Internet} censorship in {China}:
  Where does the filtering occur?'' in \emph{Passive and Active Measurement
  Conference}.\hskip 1em plus 0.5em minus 0.4em\relax Springer, 2011, pp.
  133--142.

\bibitem{Sparks2012a}
Sparks, Neo, Tank, Smith, and Dozer, ``The collateral damage of {Internet}
  censorship by {DNS} injection,'' \emph{SIGCOMM Computer Communication
  Review}, vol.~42, no.~3, pp. 21--27, 2012.

\bibitem{Fontugne2017}
\BIBentryALTinterwordspacing
R.~Fontugne, A.~Shah, and E.~Aben, ``The (thin) bridges of {AS} connectivity:
  Measuring dependency using {AS} hegemony,'' \emph{CoRR}, vol. abs/1711.02805,
  2017. [Online]. Available: \url{http://arxiv.org/abs/1711.02805}
\BIBentrySTDinterwordspacing

\bibitem{CAIDA_as_relationships_serial_2}
{Center for Applied Internet Data Analysis}, ``{AS Relationships (serial-2)}.''

\bibitem{CAIDA_as_org_info}
------, ``{Inferred AS to Organization Mapping Dataset}.''

\bibitem{RIPE_Stat}
RIPEstat, ``{RIPE Stat}.''

\bibitem{MaxMind}
{MaxMind}, ``{GeoIP2 Anonymous IP Database}.''

\bibitem{FreedomHouseInternetFreedom}
{Freedom House}, ``{Internet Freedom Scores}.''

\bibitem{Guillermo2022}
G.~Baltra and J.~Heidemann, ``What is the internet? (considering partial
  connectivity),'' University of Southern California, Tech. Rep., 2022.

\bibitem{Gill2012}
P.~Gill, M.~Schapira, and S.~Goldberg, ``Modeling on quicksand: Dealing with
  the scarcity of ground truth in interdomain routing data,'' \emph{SIGCOMM
  Comput. Commun. Rev.}, vol.~42, no.~1, p. 40–46, 2012.

\bibitem{Ensafi2015a}
R.~Ensafi, P.~Winter, A.~Mueen, and J.~R. Crandall, ``Analyzing the {Great
  Firewall} of {China} over space and time,'' \emph{Privacy Enhancing
  Technologies}, vol. 2015, no.~1, 2015.

\bibitem{Master2023a}
\BIBentryALTinterwordspacing
A.~Master and C.~Garman, ``A worldwide view of nation-state {Internet}
  censorship,'' in \emph{Free and Open Communications on the Internet}, 2023.
  [Online]. Available:
  \url{https://www.petsymposium.org/foci/2023/foci-2023-0008.pdf}
\BIBentrySTDinterwordspacing

\bibitem{Waehlisch2012}
M.~Wählisch, T.~Schmidt, M.~de~Brün, and T.~Häberlen, ``Exposing a
  nation-centric view on the german internet -- a change in perspective on the
  as level,'' in \emph{International Conference on Passive and Active Network
  Measurement}, vol. 7192, 03 2012.

\bibitem{Kohler2022}
K.~Kohler, ``{One, Two, or Two Hundred Internets? The Politics of Future
  Internet Architectures},'' \emph{{CSS Cyberdefense Reports}}, 2022.

\bibitem{rfc4272BGPVulner}
S.~L. Murphy, ``{BGP Security Vulnerabilities Analysis},'' RFC 4272, Jan. 2006.

\end{thebibliography}
\bibliographystyle{IEEEtran}

\end{document}